\documentclass[final,5p,times,onecolumn,sort&compress,hidelinks]{elsarticle}

\makeatletter
\providecommand{\doi}[1]{%
  \begingroup
    \let\bibinfo\@secondoftwo
    \urlstyle{rm}%
    \href{http://dx.doi.org/#1}{%
      doi:\discretionary{}{}{}%
      \nolinkurl{#1}%
    }%
  \endgroup
}
\makeatother 

\usepackage[normalem]{ulem}
\usepackage[utf8]{inputenc}
\usepackage{slashed}
\usepackage{graphicx}
\usepackage{bm}
\usepackage{listings}
\usepackage{pdfpages}
\usepackage{pstricks}
\usepackage{color}
\usepackage{enumerate}
\usepackage{float}
\usepackage{amsmath}
\usepackage{amssymb}
\usepackage{mathtools}
\usepackage{amssymb}
\usepackage{marvosym}
\newcommand{\be}{\begin{equation}}
\newcommand{\ee}{\end{equation}}
\newcommand{\bea}{\begin{eqnarray}}
\newcommand{\eea}{\end{eqnarray}}

\newcommand{\bfk}{\mbox{\boldmath $k$}}
\def\bkt{\bfk_\perp}

\newcommand{\bfp}{\mbox{\boldmath $p$}}

\def\bpp{\bfp_\perp}

\newcommand{\trans}[1]{\mbox{\boldmath $#1$}_\perp}

\def\masspara{\mbox{\small M}^2}

\def\massparasqrt{\mbox{\small M}}
\def\massparasqrtscript{\mbox{\scriptsize M}}
\def\power{\alpha}
\def\ptatmax{ p_{{}_{\!\!0}}{}_{\!\!{\kern 0.16em}\perp}{}}
\def\lsim{\mathrel{\rlap{\lower4pt\hbox{\hskip1pt$\sim$}}\raise1pt\hbox{$<$}}}
\def\gsim{\mathrel{\rlap{\lower4pt\hbox{\hskip1pt$\sim$}}\raise1pt\hbox{$>$}}}
\def\nostrocostruttino#1\over#2{\mathrel{\mathop{\kern 0pt \rlap
{\hbox{$#1$}}} \hbox{\kern-.135em $#2$}}}

\definecolor{dpmagenta}{rgb}{0.8, 0.0, 0.8}

\begin{document}
\begin{frontmatter}
\author{M. Boglione\fnref{label1,label2}}
\ead{boglione@to.infn.it}
\author{J. O. Gonzalez-Hernandez\fnref{label1,label2,label3}}
\ead{jogh@jlab.org}
\author{R. Taghavi\fnref{label4}}
\ead{r.taghavi@stu.yazd.ac.ir}
\address[label1]{Dipartimento di Fisica, Universit\`a di Torino, Via P. Giuria 1, 10125 Torino, Italy}
\address[label2]{INFN - Sezione Torino, Via P. Giuria 1, 10125 Torino, Italy}
\address[label3]{Department of Physics, Old Dominion University, Norfolk, VA 23529, USA}
\address[label4]{Department of Physics, Yazd University, Yazd, Iran}

\title{Transverse parton momenta in single inclusive hadron production in ${e^ + }{e^ - }$ annihilation processes.}
  
\begin{abstract}
We study the transverse momentum distributions of single inclusive hadron production in ${e^ + }{e^ - }$ annihilation processes. 
Although the only available experimental data are scarce and quite old,
%interesting effects related to the non-perturbative 
%regime of transverse momentum dependent (TMD) evolution can be observed. 
%Evidences are 
we find that the fundamental features of transverse momentum dependent (TMD) evolution, historically addressed in Drell-Yan 
processes and, more recently, 
in Semi-inclusive deep inelastic scattering processes, are visible in ${e^ + }{e^ - }$ annihilations as well.  Interesting 
effects related to its non-perturbative regime can be observed. \\
We test two different parameterizations for the $p_\perp$ dependence of the cross section: the usual Gaussian distribution and a power-law model. 
We find the latter to be more appropriate in describing this particular set of experimental data, over a relatively large range of $p_\perp$ values. 
%A discussion of how this model might be connected to the framework of TMD evolution is provided, and the caveats related to the possible 
%different interpretations of the results of our analysis are described.
We use this model to map some of the features of the data within the framework of TMD evolution, 
and discuss the caveats of this and other possible interpretations, related to the one-dimensional nature of the available experimental data.
%
%
%While waiting for more modern, high statistics experimental data on unpolarized cross sections, we develop a novel analysis of 
%the old TASSO, MARK II and PLUTO $p_\perp$ distribution measurements using 
%an up-to-date TMD factorization formalism, 
%and analysis techniques which can be straightforwardly employed for the analysis of 
%future experimental measurements. 
%to show that these data are consistent with the non-perturbative TMD evolution effects expected within a QCD factorization scheme.
%We investigate what experimental data can tell us about the non-perturbative regime of a QCD factorization scheme of TMD.
\end{abstract}

\end{frontmatter}

%%%%%%%%%%%%%%%%%%%%%%%%%%%%%%%%%%%%%%%%%%%%%%%%%%%%%%%%%%%%%%%%%%

\sloppy

%%%%%%%%%%%%%%%%%%%%%%%%%%%%%%%%%%%%%%%%%%%%%%%%%%%%%%%%%%%%%%%%%%
\section{Introduction}
\label{sec:intro}
Transverse momentum dependent distribution and fragmentation functions (TMDs) are fundamental tools to 
understand the structure of nucleons in terms of their elementary constituents, quarks and gluons.
TMDs are non-perturbative quantities which embed important correlations among partonic and hadronic  
intrinsic properties, like spin or orbital angular momentum, and their internal transverse 
motion.

TMD parton distribution functions can be interpreted, at leading twist, as number densities of partons 
carrying a light-cone momentum fraction $x$ of the parent nucleon momentum $P$.
Unpolarized and polarized TMD distribution functions have extensively been studied in Drell-Yan processes and semi-inclusive 
deep inelastic scattering (SIDIS) in the past; new-generation, dedicated experiments are currently running (like Drell-Yan at COMPASS or at RHIC) 
or are being planned (like the Electron-Ion Collider in the US and the AFTER proposal at CERN-LHC).

Of equal importance are the TMD fragmentation functions, which 
%represent the number densities of a given final hadron 
%generated by the hadronization of a parent parton. They 
embed fundamental information on the hadronization process,
%probability of producing a  
where a hadron $h$, carrying a light-cone fraction $z$ of the fragmenting parent parton, is produced. 
TMD fragmentation functions can be measured in single- 
or double-inclusive hadron production in $e^+e^-$ annihilation processes or, in a more involved way, in SIDIS, where they necessarily 
couple to a TMD distribution function. 
Even with the best SIDIS data presently available, several complications remain to be solved. Extensive recent studies can be found, 
for example, in Refs.~\cite{Signori:2013mda,Anselmino:2013lza,Aidala:2014hva,Collins:2016hqq,Boglione:2016bph}  

\begin{figure}[t]
\vspace{-1.0cm}
\includegraphics[scale=0.3]{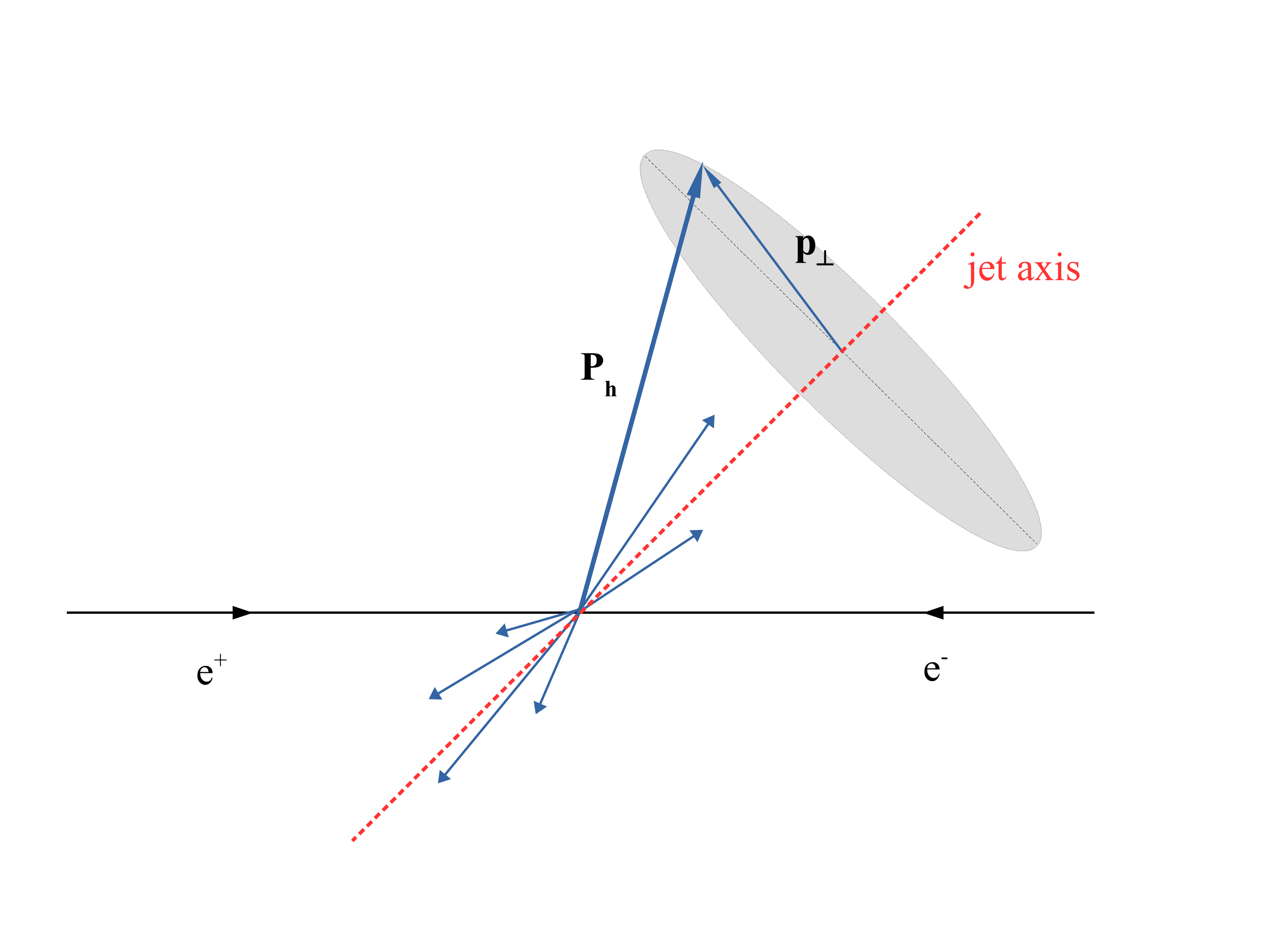} 
\vspace{-1.0cm}
\caption
{Illustration of a typical hadronic event from the $e^+e^-$ annihilation process, showing the reconstructed jet axis,
the hadron momentum $P_h$ and its transverse component $p_\perp$, perpendicular to the jet axis.
}
\label{fig:epem-jet}
\end{figure}

In $e^+e^-$ collisions, at c.m. energies below the $Z^0$ mass, the electron and positron predominantly annihilate to form a single virtual
photon, which can subsequently produce a $q\bar q$ pair. The quark and anti-quark will then convert into hadrons. At sufficiently
high energies, these multi-hadronic events are expected to form two back-to-back jets (due to the limited transverse
momentum along the original quark direction). Single inclusive distributions in variables relative to the jet
direction, which is expected to be the quark direction, will therefore give information about the fragmentation of quarks into hadrons. 
In particular, the dependence of the inclusive distributions in momentum transverse to the jet axis will provide the $golden \;channel$ 
toward the phenomenological extraction of TMD FFs. An illustration of this process is given in Figure~\ref{fig:epem-jet}.

While much effort has been put into measuring and extracting unpolarized and even polarized TMD PDFs (the Sivers function 
is a well-known example), little or no experimental information on TMD FFs is presently available. BELLE and BaBar Collaborations have 
recently presented new multi-dimensional data analyses of the Collins asymmetry in $e^+e^- \to h_1\, h_{2}\, X$ processes, which have allowed a 
first glance to the intrinsic transverse motion of partons inside hadrons through the extraction of the Collins function, a polarized, 
chirally-odd TMD fragmentation function. Relevant recent literature on this and other $e^+e^-$ related subjects are presented 
in Refs.~\cite{Kang:2014zza,Bacchetta:2015ora,Anselmino:2015sxa,Anselmino:2015fty}. 
However, no modern measurements exist of the unpolarized TMD FFs, although a thorough 
knowledge of this function would be of fundamental importance for any TMD study. 

While waiting for up-to-date, high statistics and (possibly) multidimensional results on the $p_\perp$ distribution of $e^+e^-$ unpolarized 
cross sections (the BELLE Collaboration has already presented some of their preliminary Monte Carlo simulations at SPIN 2016~\cite{Seidl:2016aa}), 
we concentrate on a set of rather old measurements 
%(data were collected between 1979 and 1986) 
of single inclusive hadron production in $e^+e^-$ annihilation processes, $e^+e^- \to h\, X$, by the TASSO Collaboration at 
PETRA (DESY)~\cite{Althoff:1983ew,Braunschweig:1990yd}: 
$p_\perp$ distributions were provided for four different c.m. energies between 14 and 44 GeV, corresponding to charged particle production 
summed over all charges and all particle species, with no flavor separation. 
Cross sections are given as functions of $p_\perp$, integrated over the energy fraction $z_h=2E_h/\sqrt{s}$ of the detected hadron $h$. 
Note that, up to corrections of order $p_\perp^2/s$, $z_h$ coincides with the light-cone momentum fraction $z$.
Although no information is offered about possible cuts applied to $z_h$, average values are provided for each c.m. energy set, 
as summarized in Table~\ref{tab1}: they correspond to rather low values, ranging from $\langle z_h \rangle$ = 0.13 at 14 Gev to 
$\langle z_h \rangle$ = 0.08 at 44 GeV.
Together with TASSO data, we also consider the analogous MARKII Collaboration measurements~\cite{Petersen:1987bq}, collected at the SLAC 
storage ring PEP, at a fixed c.m. energy of 29 GeV, and PLUTO data on the average transverse momentum square~\cite{Berger:1983yp}, 
collected at PETRA (DESY).

Crucial to all these data is the correct determination of the jet axis, to which the $p_\perp$ distributions are most sensitive, 
beside proper treatment of geometric acceptance effects, trigger bias, kinematics cuts and radiative corrections. These corrections, 
obtained by comparison to Monte Carlo simulations, are somehow model dependent. Clearly all these issues introduce very large uncertainties which, 
according to modern standards, were largely underestimated.

%%%%%%%%%%%%%%%%%%%%%%%%%%%%%%%%%%%%%%%%%%%%%%%%%%%%%%%%
%%%%%%%%%%%%%%%%%%%%%%%%%%%%%%%%%%%%%%%%%%%%%%%%%%%%%%%%

Although these data are affected by several limitations (no hadron separation, limited coverage, $z_h$ integration, 
difficulties in reconstructing the jet axis, etc), they represent an extremely interesting example of a direct 
measurement of intrinsic transverse momenta. 
In fact, as mentioned above, the jet axis resulting from each $e^+e^-$ scattering  
%the reported $p_\perp$ is then the transverse momentum of the final hadron with respect to the jet axis, see Figure~\ref{fig:epem-jet}. 
%Ideally, this axis 
identifies the direction of the fragmenting $q\bar q$ pair ($q$ and $\bar q$ should be back to back in the 
$e^+e^-$ c.m. frame if radiative effects are appropriately subtracted), 
and the detected $p_\perp$ represents a direct measurement of the transverse momentum of the final 
hadron with respect to the fragmenting parent parton, see Figure~\ref{fig:epem-jet}.
%The TASSO collaboration considered several different algorithms for the jet axis reconstruction. 
%The thrust axis is the most reliable for our purpouses. Unfortunately, TASSO published $p_\perp$ distribution are referred to 
%the sphericity axis (although no large discrepancies among different axis determinations can be observed).   

The purpose of this article is two-folded: first, we will devise and test an appropriate functional form to describe the 
$p_\perp$ distributions measured by TASSO and MARKII, achieving as much information as we possibly can on the TMD unpolarized FF; 
then, a careful interpretation of our results will be provided focusing on the features related to TMD factorization within a TMD 
evolution scheme, in the non-perturbative regime.

As these measurements offer quite limited information, we will not be able to perform a detailed extraction of the TMD 
fragmentation functions as done in the past~\cite{Anselmino:2005nn,Anselmino:2007fs,Anselmino:2012aa,Anselmino:2013lza,Anselmino:2015sxa}.
However, we will observe that interesting signatures of TMD factorization 
in the non perturbative regime can be detected in these data sets. 
In particular, we will find indications that a power-law $p_\perp$ behavior, different from the Gaussian parametrization of the TMD FFs 
we usually used in our previous analyses, might reproduce these data more successfully, especially as $p_\perp$ grows larger then a 
few hundred MeV and we enter the region in which TMD evolution effects start to become more visible. 
%
%%%%%%%%%%%%%%%%%%%%%%%%%%%%%%%%%%%%%%%%%%%%%%%%%%%%%%%%%%%
\begin{table}[t]
\centering
\begin{tabular}{|c|c|c|}
\hline
\rule{0pt}{1.2em}
%\multicolumn{2}{c}{Item} &            \\ \cline{1-2}
Experiment     & c.m. energy & $\langle z_h \rangle $ \\ 
\hline
\hline
\rule{0pt}{1.2em}
TASSO          & 14 GeV    & 0.13     \\
               & 22 GeV    & 0.11     \\
               & 35 GeV    & 0.09     \\
               & 44 GeV    & 0.08     \\ 
\hline 
\rule{0pt}{1.2em}
MARK II        & 29 GeV    & 0.09     \\ 
\hline 
\rule{0pt}{1.2em}
PLUTO          & 7.7 GeV   & --  \\
               & 9.4 GeV   & --  \\
               & 12.0 GeV  & --  \\
               & 13.0 GeV  & --  \\
               & 17.0 GeV  & --  \\
               & 22.0 GeV  & --  \\
               & 27.6 GeV  & --  \\
\hline
\end{tabular}
\caption{
Upper and central panels: center of mass energies and corresponding $z_h$ mean values 
for the TASSO and MARK II cross sections. 
Lower panel: center of mass energies corresponding to PLUTO measurements of $\langle p_\perp^2\rangle$.
}
\label{tab1}
\end{table}
%%%%%%%%%%%%%%%%%%%%%%%%%%%%%%%%%%%%%%%%%%%%%%%%%%%%%%%%%%%%

\section{Formalism}
\label{sec:formalism}

Similarly to the collinear case~\cite{Collins:2011zzd}, the ${e^ + }{e^ - }\to hX$ cross section can be casted in the following form:
\begin{equation}
\label{eq:csec}
\frac{d\sigma ^h}{dz \,d^2 \trans{p}}  = L_{\mu\nu} W^{\mu\nu} =  \frac{4\pi \alpha ^2}{3s} \,z\,F_1^h(z,p_\perp;Q^2) \,.
\end{equation}
Within TMD-factorization, up to power suppressed terms, the hadronic tensor $W$ can be expressed as
\be
\label{eq:tensorsep}
W^{\mu\nu} = W^{\mu\nu}_{TMD} + W^{\mu\nu}_{coll}  \;.
\ee
The first term on the right hand side of Eq.~(\ref{eq:tensorsep}), $W^{\mu\nu}_{TMD}$, corresponds to the region of small transverse momenta, while 
the second, $W^{\mu\nu}_{coll}$, is calculable within collinear factorization and contains corrections that become important at larger values of $p_\perp$. 
In the case at hand, as there is only one observed final hadron, one may write for the TMD term: 
\be
\label{eq:tmdterm}
W^{\mu\nu}_{TMD} \propto \sum_{f} |{\cal H}_f(Q;\mu)|^{\mu\nu} D_{h/f}(z,\bpp;\mu,\zeta_D),
\ee
where $Q$ is the hard scale of the process, $z$ and $\bpp$ are the observed hadronic variables, $\mu$ 
is the renormalization scale, while $\zeta_D$ is a regulator for the light-cone divergences that arise in TMD-factorization. 
The hard factor ${\cal H}_f$ can be calculated in perturbation theory, while
the TMD FF is defined by the relations
\be
D_{h/f}(z,z\bkt;\mu,\zeta_D)\equiv\frac{1}{(2\pi)^2}\int d^2\trans{b} e^{-i\trans{k}\cdot\,\trans{b}}\tilde{D}_{h/f}(z,\trans{b};\mu,\zeta_D)\,,
\ee
\bea
&& \hspace*{-1.5cm} \tilde{D}_{h/f}(z,\trans{b};\mu,\zeta_D)  \equiv  
\nonumber\\ && %\hspace*{-0.8cm}
\;\sum_{j}
\Bigg[\tilde{C}_{j/f}\otimes d_{h/j}(z;\mu_b)/z^2\Bigg]
\nonumber\\&&
%
%\sum_{j}\int_{z}^{1}\frac{d\hat{z}}{\hat{z}^{3-2\epsilon}}
%\tilde{C}_{j/f}(z/\hat{z},b_*)d_{h/j}(\hat{z};\mu_b)\nonumber\\&&
\hspace*{-0.3cm}
\times\,
{\rm exp}\Bigg\{\int_{\mu_b}^{\mu}\frac{d\tilde{\mu}}{\tilde{\mu}}
\Big[\gamma_D(\alpha_s(\tilde{\mu});1)-\gamma_K(\alpha_s(\tilde{\mu}))\log\left(\frac{\zeta_D}{\tilde{\mu}}\right)\Big]\Bigg\}\nonumber\\ &&
\hspace*{-0.3cm}
\times\,
{\rm exp}\Bigg\{\tilde{K}(b_*;\mu_b)\log\left(\frac{\zeta_D}{\mu_b}\right)	\Bigg\} \nonumber \\ &&
\hspace*{-0.3cm}
\times\,
{\rm exp}\Bigg\{g_{h/j}(z,b_\perp)+g_K(b_\perp)\log\left(\sqrt{\frac{\zeta_D}{\zeta_D^{(0)}}}\right)\Bigg\}\,.\hphantom{qqqqq}
\label{eq:tmdbt}
\eea
Notice the conjugate variable to $\trans{b}$ is $\trans{k}$ rather than $\trans{p}$ ($\trans{k}$ is the component of the fragmenting parton
momentum transverse to the final hadron, in a reference frame in which the latter is purely longitudinal). Details on the various ingredients 
of Eq.~\eqref{eq:tmdbt} can be found in Refs.~\cite{Collins:2011zzd,Aybat:2011zv}.
% In the above definition, $\zeta_D$ is a regulator for the light-cone 
% divergences that arise in TMD-factorization. 
For our purposes it suffices to note that 
in the first three factors of Eq.~(\ref{eq:tmdbt})
all the ingredients are calculable within perturbation theory, except for the collinear fragmentation function $d_{h/j}$.
Notice that the Wilson coefficients, $\tilde{C}_{j/f}$, 
the integral of the anomalous dimensions, $\gamma_{D}$ and $\gamma_K$, and the
the Collins-Soper (CS) kernel, $\tilde{K}$, depend on $b_\perp$ only through the quantity $b_*$, which is set
to remain smaller than some maximum $b_{max}$.
% {\color{red} Define the anomalous dimensions $\gamma_K$ and $\gamma_D$, at least in words ? }
The last factor in Eq.~(\ref{eq:tmdbt}) contains only non-perturbative information. 

There is some freedom in the definition of Eq.~(\ref{eq:tmdbt}),
encompassed in the arbitrary quantities  $\mu$, $\mu_b$, $b_*$, $\zeta_D$, $\zeta_D^{(0)}$. 
%For detailed discussions of TMD-factorization and related issues, we refer the reader to 
%Ref.~\cite{Collins:2011zzd} 
%and to the summaries in Ref.~\cite{Aybat:2011zv}. 
For our analysis we adopt the usual choices~\cite{Aybat:2011zv}, 
 $\mu\to Q$, $\zeta_D\to Q^2$ and $\mu_b=2 e^{-\gamma_E}/b_*$, where $\gamma_E$ is the 
Euler-Mascheroni constant. With these choices the CS kernel $\tilde{K}$ vanishes at order $\alpha_s$, 
so the third factor in Eq.~\eqref{eq:tmdbt} reduces to one.
To this same order, the factor containing the anomalous dimensions $\gamma_D$ and $\gamma_K$ can be expressed in a closed analytic form. 
For our purposes, it will be useful to classify the result of this integral in terms of its dependencies on $Q$ and $b_*$, namely
\bea
\label{eq:anomalousintegral}
{\rm exp}\Bigg\{\int_{\mu_b}^{\mu}\frac{d\tilde{\mu}}{\tilde{\mu}}
\Big[\gamma_D(g(\tilde{\mu});1)-\gamma_K(g(\tilde{\mu}))\log\left(\frac{\zeta_D}{\tilde{\mu}}\right)\Big]\Bigg\}\nonumber\\
\longrightarrow 
{\cal N}_\Gamma(Q)\,f_{\;\Gamma}(b_*,Q_0)\,\exp\Bigg\{\lambda_\Gamma(b_*)\log\left(\frac{Q}{Q_0}\right)\Bigg\}\,,
\eea
for which we have used the results of the Appendices in Ref.~\cite{Aidala:2014hva}.
The quantities ${\cal N}_\Gamma(Q)$, $f_{\;\Gamma}(b_*,Q_0)$ and $\lambda_\Gamma(b_*)$ are flavor independent functions that encode 
the most prominent perturbative effects in the definition of the TMD FF.
The last of these three functions, 
\be
\label{eq:lambdagamma}
\lambda_\Gamma(b_*)\equiv\frac{32}{27}\log\left(\log\frac{2 e^{-\gamma_E}}{\Lambda_{QCD}\;b_*}\right)\,,
\ee
is the most interesting
since it correlates $b_*$ with $Q$, which means it has the effect of modifying the shape of the TMD FF under evolution. 
With these considerations, one may write for the TMD FF:
\begin{align}
\label{eq:tmdbt2}
\tilde{D}_{h/f}(z,\trans{b};\mu,\zeta_D)  = &\nonumber\\
{\cal N}_\Gamma(Q)\sum_{j}
\Bigg[\tilde{C}_{j/f}&\otimes d_{h/j}(z;\mu_b)/z^2\Bigg]
e^{g_{h/j}(z,b_\perp)}
\,f_{\;\Gamma}(b_*,Q_0) \nonumber \\
\times
\exp\Bigg\{\Big(&\lambda_\Gamma(b_*)+g_K(b_\perp)\Big)\log\left(\frac{Q}{Q_0}\right)\Bigg\}\,.
\end{align}
Therefore, except for the overall 
normalization factor ${\cal N}_\Gamma(Q)$, the effects of evolution at order $\alpha_s$ can be mapped to either 
the non-perturbative function $g_K(b_\perp)$, or the perturbative quantity $\lambda_\Gamma(b_*)$.

In the region where TMD effects dominate (see Eqs.~\eqref{eq:csec}-\eqref{eq:tmdterm}), flavor independence
of $g_K$ and $\lambda_\Gamma$ in Eq.~\eqref{eq:tmdbt2} implies
\begin{align}
\label{eq:invfourier}
{\cal F}^{-1}\left\{\frac{d\sigma ^h}{dz \,d^2 \trans{p}} \right\}\propto\,
&\exp\Bigg\{\Big(\lambda_\Gamma(b_*)+g_K(b_\perp)\Big)\log\left(\frac{Q}{Q_0}\right)\Bigg\}\,\Bigg|_{b_\perp\to z\; b_\perp}\,,
\end{align}
where the symbol ${\cal F}^{-1}$ indicates the two-dimensional inverse Fourier transform, from momentum to impact parameter space, and the 
transformation $b_\perp\to z \;b_\perp$ is needed to account for the extra factor of $z$ that appears in the definition of 
Eq.~\eqref{eq:tmdterm}, compared to the TMD term in the hadronic tensor in Eq.~\eqref{eq:tensorsep}. In the following 
section we will use relations~\eqref{eq:lambdagamma}-\eqref{eq:invfourier} to make an interpretation of our results.

\section{Data fitting and results}
\label{sec:data}

A full analysis within a TMD-evolution scheme should include all of the contributions in Eq.~\eqref{eq:tensorsep}, 
as well as a matching prescription to interpolate between regions of small and large $p_\perp$, as originally prescribed in Ref.~\cite{Collins:1984kg}.
It is important to stress that all of these ingredients provide crucial 
constraints that any full analysis should include. Some recent studies related to the complications involved in the matching prescriptions for SIDIS 
processes are presented in Refs.~\cite{Boglione:2014oea,Collins:2016hqq}. 
However, for such type of analyses, multidimensional data sets are most suitable, where
one can completely disentangle the effects of different kinematics variables.
In the case of the measurements of Ref.~\cite{Althoff:1983ew,Braunschweig:1990yd,Petersen:1987bq,Berger:1983yp}, 
the large systematic uncertainties induced by $z$-integration can only render limited insight on TMD-effects. 
However, it is still interesting to investigate what 
information about evolution can be extracted from these measurements. In fact, even for these $z$-integrated cross sections, one may expect the
{\it shape} of the $p_\perp$ distributions to be affected by TMD-evolution effects.

In order to address this question it becomes essential to make an estimate of where the large transverse momentum corrections
start becoming important in the TASSO and MARKII data sets.
To do so, we start by considering the errors of the TMD approximation, which are of order ${\cal O}(k_\perp/ Q)= {\cal O}(p_\perp/ z Q)$. 
In general, one expects that at $p_\perp\sim z Q$, the cross section should receive contributions 
from the collinear term in Eq.~\eqref{eq:tensorsep}.
Using the average values of Table~\ref{tab1}, one may estimate that this contributions should be significant at 
$p_\perp\sim2\,{\rm GeV}$. 
We identify this as the matching region.
In this article we will attempt to extract information only about the non-perturbative evolution carried by $g_K$ in 
Eq.~(\ref{eq:tmdbt2}), so we will constrain our analysis to the region of $p_\perp < 1.0\;{\rm GeV}$. 
Note that we also impose a lower cut on $p_\perp$, such that $p_\perp > 0.03$ GeV; this amounts to excluding the first data point 
in the TASSO data sets.

% We warn the reader that while $g_K$ is strongly universal, in phenomenological studies some of the perturbative 
% effects may in practice be ``absorbed'' by the parametric forms adopted for this function. Thus, one should be careful
% to distinguish the universal function $g_K$ from quantities extracted from fits, as pointed out in Ref.~\cite{Collins:2014jpa}). 
%

In order to relate the shape of the data to possible TMD-evolution effects, one needs a model
that can reproduce the transverse momentum distribution of the final hadron $h$. 
Then, by looking at the $b_\perp$-space, one may connect the parameters of the model to some of the information contained in the 
definition of Eq.~(\ref{eq:tmdbt}), as discussed in Ref.~\cite{Aidala:2014hva}.

% In the present analysis we will mostly look at the asymptotic behavior of $g_K$, so that our reference will be the 
% limiting cases of Eqs.~(\ref{eq:tmdbtasy}-\ref{eq:tmdktsmallsimple}).

At least two functional forms have been shown to appropriately describe transverse momentum distributions~\cite{Aidala:2014hva,Anselmino:2013lza}.
\begin{itemize}
 \item \underline{Gaussian form}: it is the most commonly used parametrization for phenomenological studies, 
 it has been shown to reproduce experimental data very successfully in both Drell-Yan and SIDIS processes. 
 It has the advantage of being easy to integrate analytically. 
 \item \underline{Power-law}: it is very flexible, even with a limited number of free parameters. 
 Not only can it appropriately reproduce the behavior of the cross section at small $p_\perp$, but it can also incorporate its
 tail at larger $p_\perp$ values.
\end{itemize}
We will consider both of these functional forms in order to describe the low-$p_\perp$ region of the data.
Our aim is to focus on the kinematics ranges where TMD-effects are dominant. 
As discussed above, one may estimate that perturbative effects will start to become important roughly 
around $p_\perp\sim 2\,{\rm GeV}$, but could in fact be non-negligible at even smaller
values of transverse momentum, especially as the data we consider are integrated over $z$. 
Our working hypothesis is that for $p_\perp<1\;{\rm GeV}$ the TMD-term in Eq.~(\ref{eq:tensorsep}) is the largest contribution
to the cross section.

We model the structure function $F_1$ in Eq.~(\ref{eq:csec}) so that
\bea
\frac{d\sigma ^h}{dz\,d^2\trans{p}} &=& \frac{4\pi \alpha ^2}{3s}\sum\limits_q {{e_q}^2}\, D_q^h(z,{p_\perp};{Q^2}) \nonumber \\ 
%+ D_{\bar q}^h(z,{p_\perp};{Q^2}),
&=&\frac{4\pi \alpha ^2}{3s}\sum\limits_q {{e_q}^2}\, D_q^h(z,Q^2) \;h(p_\perp)\,,
\label{eq:F1model}
\eea
which extends the leading order expression for the collinear cross section.
In Eq.~\eqref{eq:F1model}, the sum runs over all $q$ and $\bar q$ flavors, and we have assumed that the TMD fragmentation function may be written as
\be
\label{eq:tmdmodel}
D_q^h(z,{p_\perp})= D_q^h(z) \;h(p_\perp)\,,
\ee
where $D_q^h(z)$ is the collinear, unpolarized FF (which we take from Ref.~\cite{deFlorian:2007hc}). 
The function $h(p_\perp)$ incorporates all of the $p_\perp$ dependence of the TMD FF, it is flavor independent and it is 
normalized so that it integrates to unity.

The TASSO collaboration provides cross sections differential in $p_\perp$, normalized to the fully inclusive cross section, 
which at leading order read
\be
\sigma_{tot} \overset{LO}{=} \sigma_{0}=\frac{4\pi \alpha ^2}{3s}\,\sum_q {{e_q}^2},
\label{eq:sigmatotlo}
\ee
so that our fits will involve the expression
\be
\frac{1}{\sigma_0}\frac{d\sigma ^h}{dp_\perp} = 2 \pi p_\perp\left[\int \!\! dz\,\frac{\sum_q e_q^2\, D_q^h(z;Q^2)}{\sum_q e_q^2}\right]h(p_\perp)\,. 
\label{eq:csecfit}
\ee
A note of caution is necessary at this point. 
The TASSO distributions in $p_\perp$ at different energies cannot be described by simply using the model of Eq.~(\ref{eq:csecfit}). 
Instead, one must incorporate a treatment for their normalizations at different values of $Q$. While this
may be in conflict with a possible probabilistic interpretation of the function $h(p_\perp)$,
it should not affect our conclusions regarding TMD evolution, since for that, we will focus on aspects of the $p_\perp$ distributions that
regard their shape, but not the precise values of their maxima. We note that accounting for all the features of these data may be
challenging even within a full TMD analysis, as large systematic errors may translate into out-of-control 
normalizations when dealing with $z$-integrated data.

The fits in the next subsections are performed on the TASSO $p_\perp$-distributions \emph{only}. We will use 
the MARK II $p_\perp$-dependent normalized cross section and the PLUTO measurements of $\langle p_\perp^2 \rangle$
to cross-check our results. When appropriate, error bands corresponding to a $2\sigma$ confidence level 
are provided, obtained by generating random points in the parameter space for which
$\chi_i^2\in [\chi_0^2,\chi_0^2+\Delta \chi^2]$, where $\chi_0^2$ is the minimal value given by the fit
and $\Delta \chi^2$ depends on the number of parameters of the model; the relevant cases in our fits 
involve either $6$ or $7$ free parameters, which correspond to $\Delta \chi^2$ values of $12.85$ and $14.34$, respectively.

\subsection{Gaussian shape at low $p_\perp$ \label{sec:gauss}}

\begin{figure}[t]
\centering
 \includegraphics[scale=0.7]{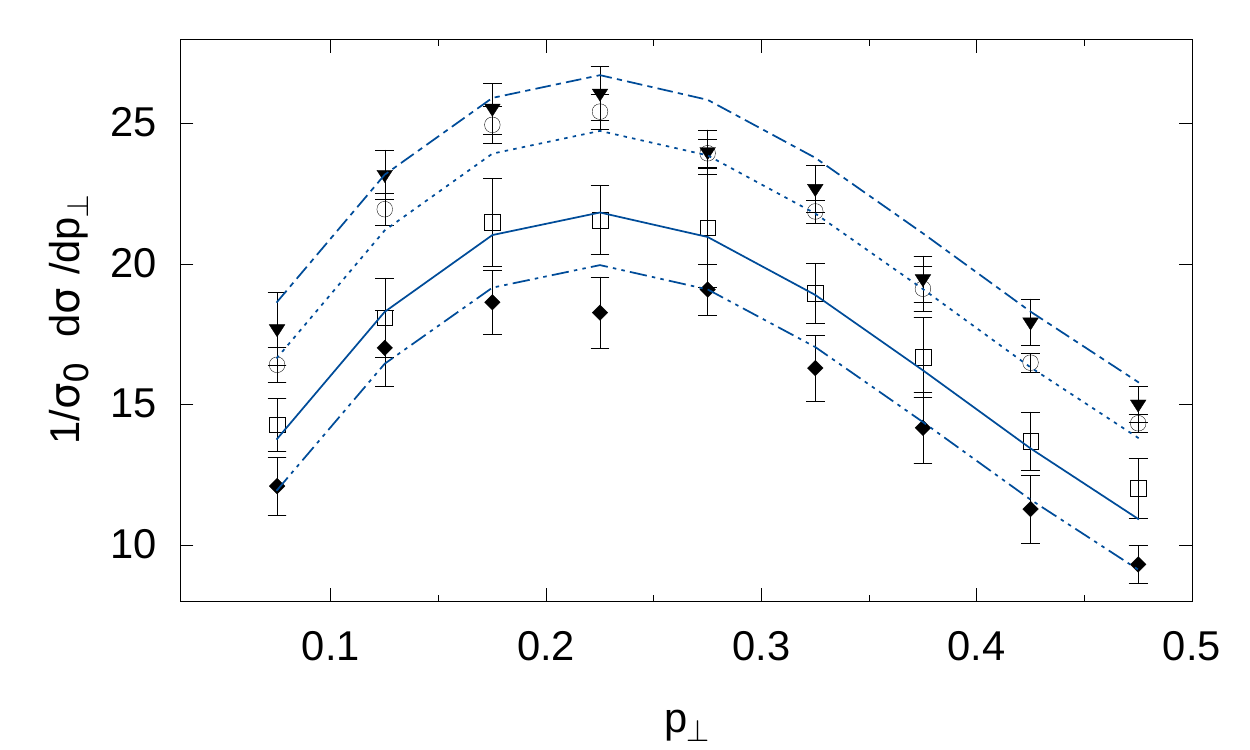} 
 \includegraphics[scale=0.7]{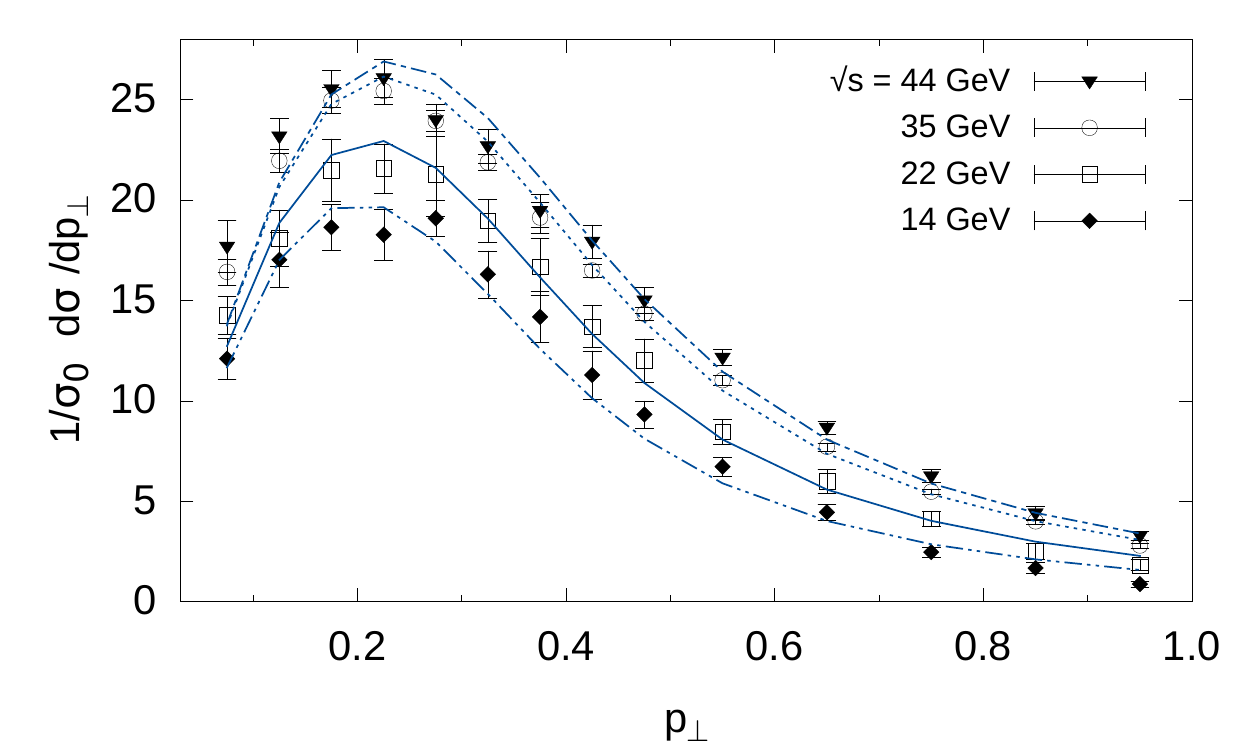} 
\caption
{
  Gaussian description of the TASSO $p_\perp$ distributions at 4 different c.m. energies~\cite{Braunschweig:1990yd}. 
  On the upper panel the Gaussian model with constant width, see Eqs.~\eqref{eq:gauss} and \eqref{eq:shift}, up to $p_\perp = 0.5$ GeV. 
  On the lower panel the Gaussian model with the Q dependent width of Eq.~\eqref{eq:widthQdep}, up to $p_\perp = 1.0$ GeV.  
}
\label{fig:gaussian-model}
\end{figure}

\noindent
We start by applying a Gaussian model with a constant width:
\be
h(p_\perp)=\frac{{e^ {- {p_\perp}^2/\langle {{p_\perp}^2} \rangle }}}{{\pi \langle {{p_\perp}^2} \rangle }}\,.
\label{eq:gauss}
\ee
As mentioned earlier, in order to describe the data we need an appropriate 
treatment for the normalization. Unexpectedly, we find that using different multiplicative constants, one for each energy, is not enough
to obtain a good fit, even in the limited region of $p_\perp<0.5$ GeV (see the first entry in Table~\ref{tab:gaussian-fit}).
Instead, by also introducing a $Q$-dependent shift for the cross sections, so that one has
\be
N\;\frac{1}{\sigma_0}\frac{{d{\sigma ^h}}}{{dz\,{d^2}{{\bf p}_\perp}}} +\delta \,Q\,,
\label{eq:shift}
\ee
one can obtain a good description of the data, as seen in the second entry of Table~\ref{tab:gaussian-fit} and on the upper
panel of Figure~\ref{fig:gaussian-model}.
This unconventional prescription to deal with the normalization, while leaving little room for a partonic 
interpretation for the function $h(p_\perp)$, allows us to verify quantitatively that as far as the width of the
$p_\perp$ distribution is concerned, no significant change can be observed with growing Q. One may not conclude, however,
that TMD-effects do not appear in these transverse momentum ranges, but rather that the data analyzed do not 
have the necessary accuracy to show possible width changes in this region. Thus, one must 
try to extend the description of the data to larger values of $p_\perp$.

%%%%%%%%%%%%%%%%%%%%%%%%%%%%%%%%%%%%%%%%%%%%%%%%%%%%%%%%%
%%%%%%%%%%%%%%%%%%%%%%%%%%%%%%%%%%%%%%%%%%%%%%%%%%

%%%%%%%%%%%%%%%%%%%%%%%%%%%%%%%%%%%%%%%%%%%%%%%%%%%%%%%%%%%%%%%%%%%%%%%%%%%%%%%
\begin{table*}[ht]
 \centering
\begin{tabular}{|c|c|l|r|}
  \hline
  \hline
\rule{0pt}{1.2em}
Parametrization & Normalization & ~~~~~~~~~ Gaussian width  & $\chi^2_{pt}$ \rule[-0.5em]{0pt}{1.0em}\\ 
\hline
\hline
\rule{0pt}{1.2em}
Gaussian - I\,\,\,\,\,     & $N=\{N_{14},N_{22},N_{35},N_{44}\}$                    & $\langle p_\perp ^2 \rangle =$  constant              &      \rule[-1.0em]{0pt}{1.0em}  \\
$p_\perp\in[0.03 - 0.50]$ GeV& $N_{14}=2.3\pm 0.2$,  \,\,\, $N_{22}=2.7\pm 0.2$        & $\langle p_\perp ^2 \rangle = 0.118 \pm 0.004$ GeV$^2$ &    5.9                 \rule[-0.5em]{0pt}{1.0em}  \\
           36 data points                & $N_{35}=3.1\pm 0.1$,  \,\,\, $N_{44}=3.2\pm 0.1$       &                                                         &                            \rule[-0.5em]{0pt}{1.0em}  \\  
\hline
\rule{0pt}{1.2em}
Gaussian - II \,\,      & $N$, $\delta \, Q$     & $\langle p_\perp ^2 \rangle = $ constant                  &     \rule[-1.0em]{0pt}{1.0em}  \\
$p_\perp\in[0.03 - 0.50]$ GeV& $N=1.8\pm 0.2$         & $\langle p_\perp ^2 \rangle = 0.098 \pm 0.005$ GeV$^2$     &   0.74   \rule[-0.5em]{0pt}{1.0em}  \\
          36 data points                  & $\delta=0.22\pm 0.03$    &                                                          &      \rule[-0.5em]{0pt}{1.0em}  \\
\hline
\rule{0pt}{1.2em}
Gaussian - III\,       & $N=\{N_{14},N_{22},N_{35},N_{44}\}$ & $\langle p_\perp ^2 \rangle = 2g_1 + 2g_2 z^2 \log\frac{Q}{3.2}$ &     \rule[-1.0em]{0pt}{1.0em} \\
$p_\perp\in[0.03 - 1.00]$ GeV  & $N_{14}=2.7\pm 0.2$, \,\,\,$N_{22}=3.3\pm 0.3$        &    $g_1= 0.013 \pm 0.004$ GeV$^2$                &    2.7                 \rule[-0.5em]{0pt}{1.0em}  \\
            56 data points                   & $N_{35}=4.0\pm 0.1$, \,\,\,$N_{44}=4.3\pm 0.2$        &    $g_2= 2.6 \pm 0.3$ GeV$^2$                    &                     \rule[-0.5em]{0pt}{1.0em}  \\ 
\hline  \hline
\end{tabular}
\caption{Fits of the TASSO four sets of cross sections, corresponding to the Gaussian parameterization of the $p_\perp$ distributions. 
Parametrization I refers to the usual choice of Eq.~\eqref{eq:gauss}, Parametrization II refers to the Gaussian corrected by a 
Q-dependent shift, see Eq.~\eqref{eq:shift}, while Parametrization III corresponds to a Gaussian distribution with a Q dependent width, 
as in Eq.~\eqref{eq:widthQdep}.}
\label{tab:gaussian-fit}
\end{table*}
%%%%%%%%%%%%%%%%%%%%%%%%%%%%%%%%%%%%%%%%%%%%%%%%%%%%%%%%%%%%%%%%%%%%%%%%%%%%%%%%%%%%%%%%%%%%%%%%%%%%%%%%%%%%%%%%%%%%%%%%%%%%%%%%%%%%%%%5

%  Trying to extend our description to larger $p_\perp$ values will result in a dramatic increase of the $\chi^2$.
%  We have tested different ways to accommodate the proper $Q$ dependence on the normalization of the data. 
%  For instance, if we introduce independent normalizations for each c.m. energy, the Gaussian model can 
%  (to some extent) reproduce the onset of the tail of the cross 
%  sections up to $p_\perp = 1$ GeV, provided we allow the width $\langle p_\perp^2 \rangle$ to also vary (evolve?) with $Q$. 
%  

For $p_\perp>0.5\,{\rm GeV}$, a noticeable dependence of the distributions on the c.m energy suggests that  
a $Q$-dependent width may be appropriate. However, even with this extension of the model, describing the data
past this point turns out to be extremely difficult. To illustrate this, we consider the functional form 
 \be
 \label{eq:widthQdep}
 \langle p_\perp^2 \rangle = 2\,g_1+2\,g_{2}\;z^2\log\left(\frac{Q}{2\,Q_0}\right)\,,
 \ee
 which allows for some comparison to previous phenomenological studies~\cite{Landry:2002ix,Konychev:2005iy}
 where, within a CSS evolution scheme, a Gaussian behavior of the non-perturbative 
 Sudakov factor in the $b_\perp$ space was assumed.
 This time, we use a multiplicative constant for each value of $Q$ ($\delta=0$ in Eq.~\eqref{eq:shift}), and 
 fix $Q_0=1.6\,{\rm GeV}$. 
 The last entry of Table~\ref{tab:gaussian-fit} shows the results of the fit for $p_\perp<1\,{\rm GeV}$, using
 the extended Gaussian model corresponding to Eq.~(\ref{eq:widthQdep}). The obtained minimal $\chi^2$ value points
 to a rather low quality of the description of the data. The corresponding plot, 
 in the lower panel of Figure~\ref{fig:gaussian-model}, shows that there is some tension between a 
 successful description of the peak at low $p_\perp$ and an equally good description of the tail 
 at larger $p_\perp $ values. As we will show in the next subsection, the power-law can in fact accommodate for both 
 of these features of the data.
%  Fig~\ref{fig:gaussian-model} shows that the Gaussian shape fails to describe the onset of the cross section tail even including the shift 
%  to fix the normalization. 

\subsection{Power-law shape at low and moderate $p_\perp$ \label{sec:powerlaw}}

So far, we have tried the Gaussian ansatz for the $p_\perp$ distributions. We have found that 
the data favors a constant width, for values up to $p_\perp=0.5\,{\rm GeV}$. However, this can only be achieved 
by introducing a $Q$-dependent shift, which cannot be easily interpreted within a partonic picture. The Gaussian
class of models does not seem to be appropriate for larger values of $p_\perp$. 

In order to describe the data up to $p_\perp=1\,{\rm GeV}$, we test a power-law parametrization, given by
\be	
\label{eq:powerlaw}
h(p_\perp)=2(\power-1)\massparasqrt^{2\;(\power-1)}\frac{1}{\left(p_\perp^2+\masspara\right)^\power}\,,
\ee
where the factor $2(\power-1)\massparasqrt^{2\;(\power-1)}$
%in front of the power-law 
is set so that $h(p_\perp)$ integrates
to unity. The two-dimensional inverse Fourier transform of this function has an exponentially decaying asymptotic behavior, consistent with what would be expected 
from general arguments within quantum field theory, as discussed in Refs.~\cite{Collins:2011zzd,Collins:2014jpa}.
%, of the form \(b_\perp^p \, e^{-\mu b_\perp}\). 
In impact parameter space the power-law becomes
\begin{align}
\label{eq:besselK}
{\cal F}^{-1}&\left\{  
\frac{1}{\left(p_\perp^2+\masspara\right)^\power}
\right\}
=
\frac{1  }{2^\power\, \pi\,  \Gamma (\power)}\left(\frac{b_\perp}{\massparasqrt}\right)^{\power-1}\!\!\!
K_{1-\power}(b_\perp \massparasqrt) \nonumber \\
&\overset{{\rm large}\,b_\perp}{\longrightarrow}
\frac{1  }{2^\power\, \pi\,  \Gamma (\power)}\left(\frac{b_\perp}{\massparasqrt}\right)^{\power-1}\!\!\!
\sqrt{\frac{\pi}{2}}\frac{e^{-b_\perp \massparasqrtscript}}{\sqrt{b_\perp \massparasqrt}} 
\left[1+{\cal O}\left(\frac{1}{b_\perp\massparasqrt}\right)\right]\,,
\end{align}
where $K_{1-\power}(b_\perp \massparasqrt)$ is the modified Bessel function of the second kind.
In what follows, we use an independent normalization for each value of $Q$.
As discussed before, we are interested in the shape of the distributions, rather than
on their overall normalizations.

In the power-law parametrization of the $p_\perp$-differential cross section, the parameter $\masspara$ is related to the 
position of its peak, $\ptatmax$, 
by the relation $\masspara=\,\big(2\,\alpha-1\big)\ptatmax^{\!2}$.
Since the studied distributions reach their maximum at roughly the same value of transverse momentum, 
$p_\perp\approx 0.212\,{\rm GeV}$, for all values of the c.m. energy, 
in our main analysis we have imposed the conditions that
\begin{align}
\label{eq:mupconstraint}
\masspara=&\,\big(2\,\alpha-1\big)\ptatmax^{\!2} \nonumber\\
\ptatmax=& \,0.212\,{\rm GeV}\,.
\end{align}
%where $\ptatmax$ is the value of $p_\perp$ at which the distribution peaks. 
We have verified that setting $\masspara$ free, 
does in fact satisfy Eqs.~(\ref{eq:mupconstraint})
within errors. It is, however, useful to reduce the number of parameters by directly imposing these relations.

% The key feature to study in this case is the rate at which the distributions fall to zero, encoded in the 
% parameter $\power$ of Eq.~\eqref{eq:powerlaw}. 
% A very distinctive signature of TMD evolution is that it broadens
% the TMD FF as $Q$ increases (see for instance Figure~4 in Ref.~\cite{Aybat:2011zv}). 

\begin{figure}[t]
\centering
 \includegraphics[scale=0.7]{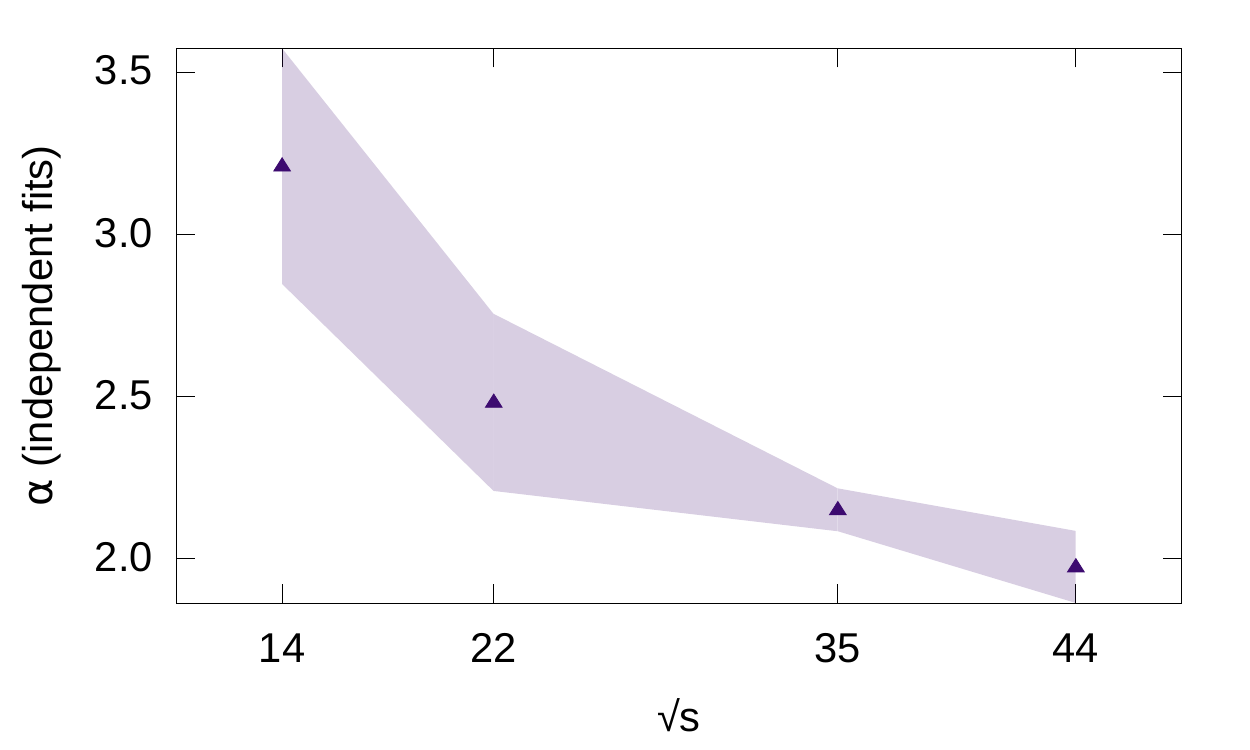} 
\caption
{
  Values of $\power$ in Eq.~\eqref{eq:powerlaw} that best describe the distributions 
  of~\cite{Braunschweig:1990yd}. The triangles represent the minimal values of independent
  fits, performed for each value of the center of the c.m energy $\sqrt{s}=Q$. 
  The shaded bands indicate the corresponding uncertainty.
}
\label{fig:pvsQ}
\end{figure}

\begin{figure}[t]
\centering
% \begin{tabular}{c@{\hspace*{1mm}}c@{\hspace*{1mm}}c}
 \includegraphics[scale=0.7]{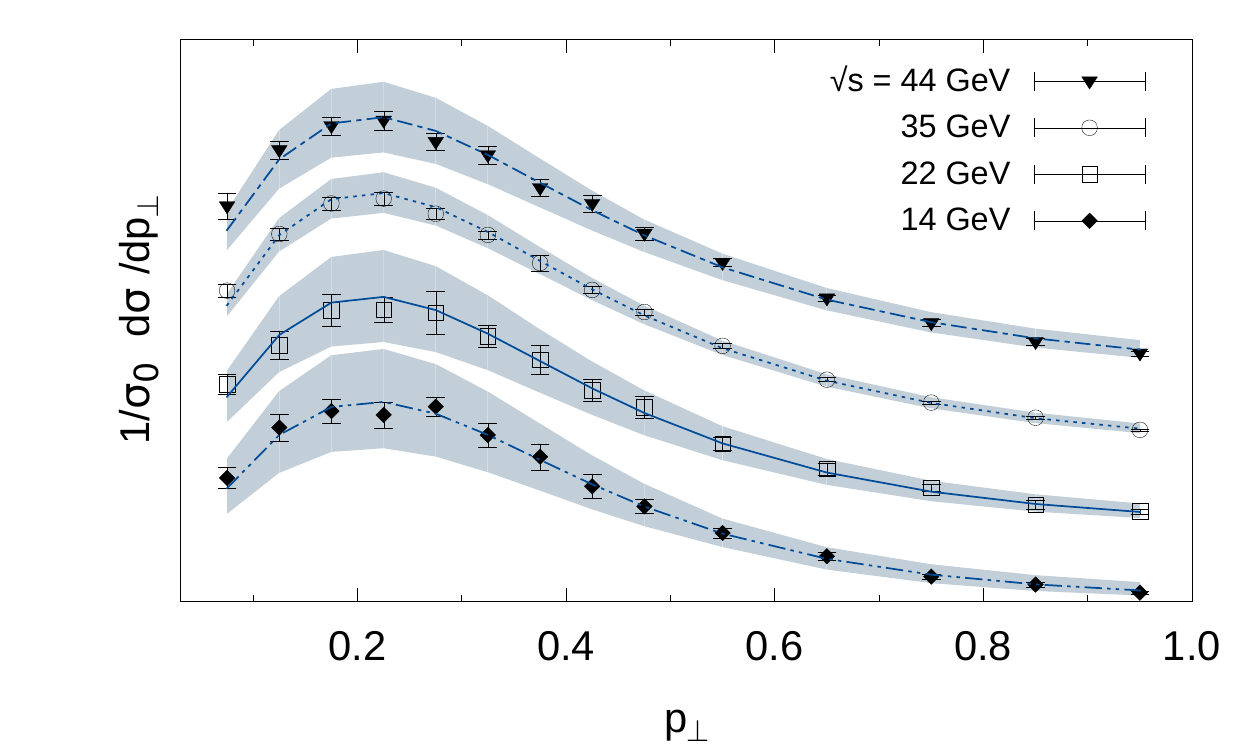} 
    
    \includegraphics[scale=0.7]{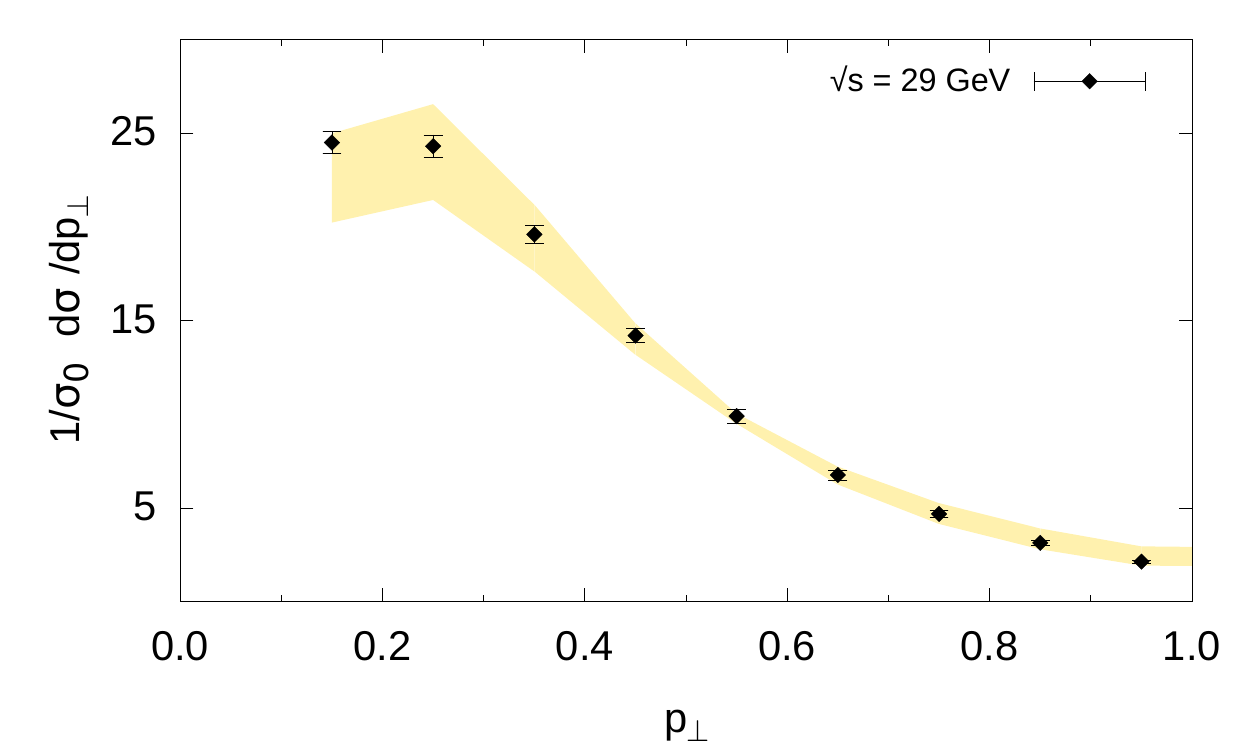}
%   \\%[3mm]
%    (a)  (b)
%    \\
%    \includegraphics[scale=0.7]{power5_Nlog_003-400_tasso}
%   & 
%    \includegraphics[scale=0.7]{power5_Nlog_003-550_tasso} 
%  \\%[3mm]
%   (e) & (f) 
% \end{tabular}
\caption
{
  In the top panel we show the results from our fit to TASSO experimental data using 
  the power-law of Eqs.~\eqref{eq:powerlaw} and~\eqref{eq:powerlog},
  in the range $0.03\,{\rm GeV}<p_\perp<1.0\,{\rm GeV}$. 
  To avoid overlapping and provide a clear display of the $2\sigma$ error bands,
  we plot the distributions for different energies applying 
  an arbitrary shift.
  In the bottom panel we compare the results from the fit to TASSO data to the MARK II cross section. 
  Note that a different normalization and its uncertainties have to be determined independently 
  for this data set. We dont display the first bin, centered at $p_\perp=0.05$ GeV.
}
\label{fig:tassofit}
\end{figure}

First, to test that the power-law can appropriately describe the data,
we conducted simple independent fits for each value of the c.m. energy, which renders one value of $\power$
for each $Q$. The most interesting aspect
of this preliminary fit is that it shows a clear dependence of the parameter $\power$ on $Q$, despite the use
of independent normalizations.
The trend of the values for $\power$ is displayed in Figure~\ref{fig:pvsQ}, which shows a decrease of the its 
optimal value with $Q$.
Due to the large uncertainties in the determination of $\power$, there are
likely several functional forms that can accommodate the observed behavior.
In order to make an educated guess for a suitable $Q$-dependence in $\power$, we 
assume that the integration over $z$ does not alter the structure of the relation in Eq.~\eqref{eq:invfourier}, 
namely
\begin{align}
\label{eq:invfourier2}
{\cal F}^{-1}\left\{\frac{d\sigma ^h}{d^{\,2} \trans{p}} \right\}\propto
&\exp\Bigg\{\tilde{g}(b_\perp)\log\left(\frac{Q}{Q_0}\right)\Bigg\}\,\,,
\end{align}
for some function $\tilde{g}(b_\perp)$. Thus, one can see that a logarithmic behavior for $\power$ may be
appropriate by 
looking at the asymptotic limit of the power-law in $b_\perp$-space, Eq.~\eqref{eq:besselK}. 
First, for the values $\power_0$ and $\power$
that describe the data at two given values $Q_0$ and $Q$, if \eqref{eq:invfourier2} holds, one should have
\begin{align}
\label{eq:relationpowers}
b_\perp^{\,\power_0}\,
\exp\Bigg\{\tilde{g}(b_\perp)\log\left(\frac{Q}{Q_0}\right)\Bigg\}
\propto
b_\perp^{\,\power}\,,
\end{align}
in the large-$b_\perp$ limit. This can be achieved if 
\begin{align}
\label{eq:gtildebehavior}
\tilde{g}(b_\perp) \overset{{\rm large}\,\,b_\perp}{\longrightarrow} \tilde{\alpha} \log(\nu\,b_\perp)\,,
\end{align}
for some values $\tilde{\alpha}$ and $\nu$, which in turn provides the relation
\begin{align}
\label{eq:powerlog}
\power  = \power_0 + \tilde{\power}\log\left(\frac{Q}{Q_0}\right)\,.
\end{align}
We have implemented Eq.~\eqref{eq:powerlog} into a fit and confirmed that in fact it reproduces
well the TASSO data. Results are shown on the third panel of Table~\ref{tab:power-law-fit}, and in
the top plot in Figure~\ref{fig:tassofit}, which includes errors corresponding to a $2\sigma$ 
confidence level. We compare these results to the MARK II data set in the bottom panel of Figure~\ref{fig:tassofit}.

The argument presented above, leading to a logarithmic $Q$-dependence for the power $\power$, 
has some caveats. First, it depends on whether Eq.~\eqref{eq:invfourier2}
is approximately correct. Furthermore since it considers only the asymptotic large-$b_\perp$
behavior of Eq.~\eqref{eq:besselK}, Eq.~\eqref{eq:powerlog} does not need to hold
for values larger than $p_\perp\sim \massparasqrt$. Therefore, even if one can describe the data,
any interpretation of the logarithmic behavior of $\power$, in terms of the ingredients
that define the TMD FF, Eq.~\eqref{eq:tmdbt}, should be taken with great caution. Nonetheless,
it is worthwhile to explore this possibility. 

Notice that the function $\tilde{g}$ acquires its behavior from $\lambda_\Gamma$ and $g_K$. Since 
the first varies slowly with $b_\perp$, and in fact does freeze to a constant value at large enough $b_\perp$ 
(see Eq.~\eqref{eq:lambdagamma}), one may see Eq.~\eqref{eq:powerlog} 
as the manifestation of a logarithmic large-$b_\perp$ trend for $g_k(b_\perp)$, analogous
to Eq.~\eqref{eq:gtildebehavior}. A behavior consistent with discussions in Refs.~\cite{Aidala:2014hva,Collins:2014jpa}.
%Of course, 
%However, this argument is not completely conclusive since one may well reproduce the data by different
%assumptions. 
However, we stress that one may well reproduce the data by using different assumptions. 
As a counter example we consider a simple picture where the cross section takes the form 
\be
\frac{d\sigma}{dz\,d^2\trans{p}}\propto\left(\frac{1}{p_\perp^2+z\,\tilde{\massparasqrt}^2}\right)^{\,\beta_1+\,\beta_2\,z}\,,
\ee
and where one accounts for the integration over $z$ by some average value $\langle z\rangle$, leading to 
\be	
\label{eq:powerlawz}
\frac{d\sigma}{d^2\trans{p}}\propto\left(\frac{1}{p_\perp^2+\langle z\rangle\,
\tilde{\massparasqrt}^2}\right)^{\,\beta_1+\,\beta_2\,\langle z\rangle}\,,
\ee
where the parameters $\beta_1$, $\beta_2$ and $\tilde{\massparasqrt}$ are to be determined by a fit. 
In this case, it is indeed possible to obtain a good description of the data by using the experimental
average values of $z$ of Table~\ref{tab1}, since they exhibit a seemingly logarithmic trend. 
However, it is difficult to make a connection to TMD evolution
since the values in Table~\ref{tab1} are in general affected by 
correlations between $Q$ and $z$ of different origin. In fact, notice that the condition of Eq.~\eqref{eq:mupconstraint}
implies a logarithmic behavior for our fit parameter $\masspara$. It is possible that, for instance, 
the effects of the TMD evolution, encoded in $g_K$, result in changes in the power $\power$ that fit the data,
while the changes in $\masspara$ are the result of correlations induced by the integration over $z$.
It seems, however, that the lack of information about the $z$-dependence of the TMD FF in the TASSO and MARK II measurements 
hinders a more solid conclusion about TMD evolution effects in these data sets.

\begin{figure}[t]
\centering
 \includegraphics[scale=0.7]{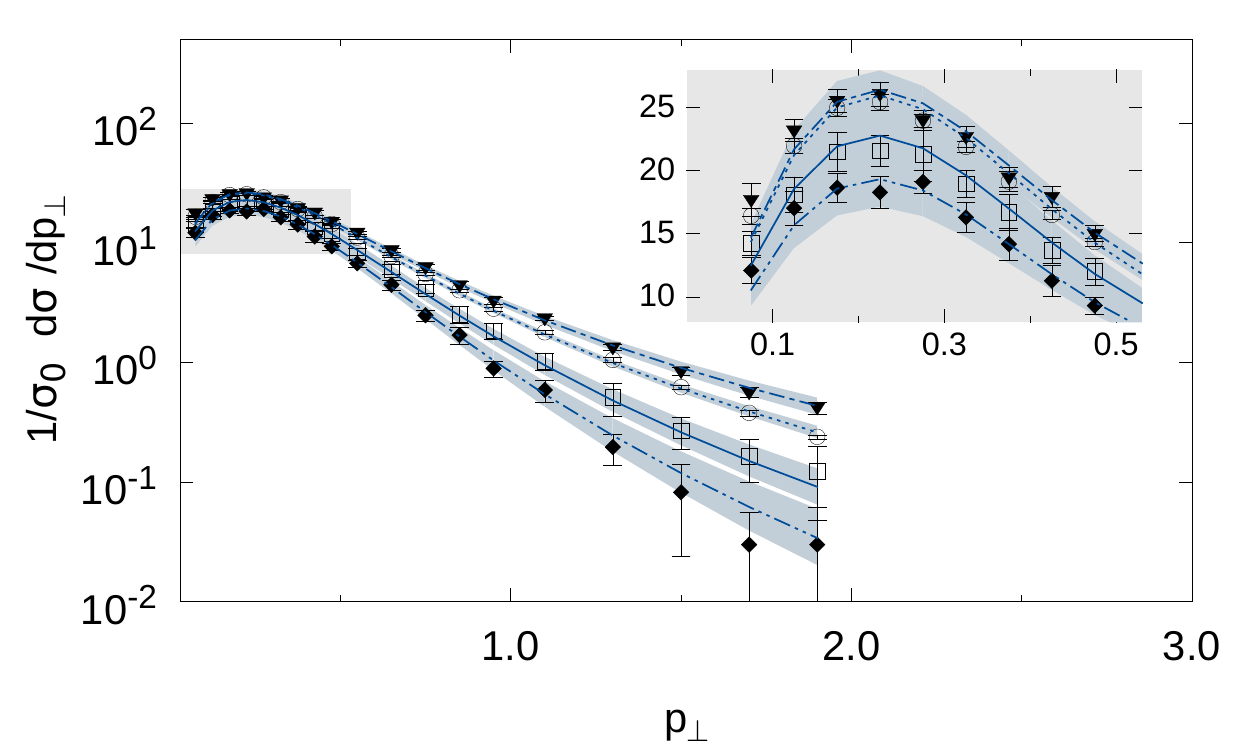} 
\caption
{Results obtained by using the power-law of Eqs.~\eqref{eq:powerlaw} and~\eqref{eq:powerlog} compared to 
TASSO $p_\perp$-dependent distributions~\cite{Braunschweig:1990yd}, in the range 
$0.03\,{\rm GeV}<p_\perp<2.0\,{\rm GeV}$. 
Error bands are computed using a $2\sigma$-confidence level, as explained in Section~\ref{sec:data}.
}
\label{fig:tassofit2}
\end{figure}

\begin{figure}[t]
\centering
 \includegraphics[scale=0.7]{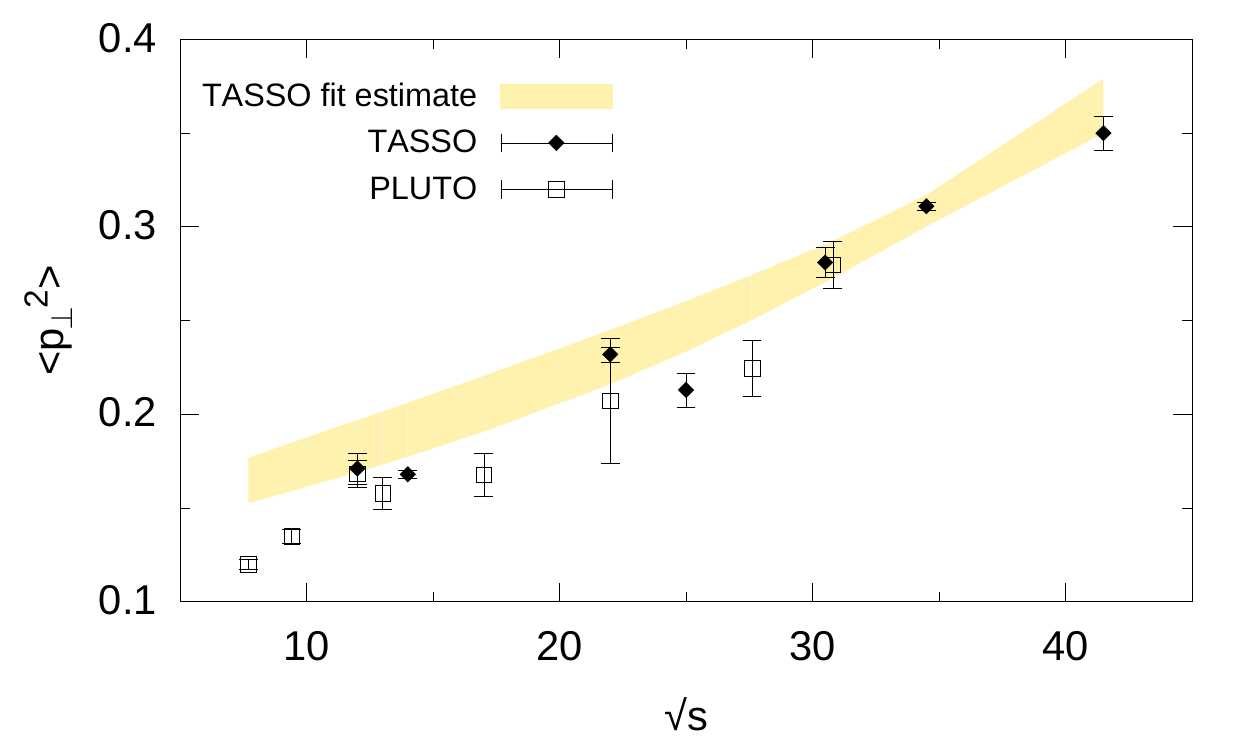} 
\caption
{Estimation of the transverse momentum mean value, $\langle p_\perp^2 \rangle$, obtained by
 using the parameters extracted by fitting the TASSO $p_\perp$ distributions up to 3.0 GeV. 
 Empty squares correspond to PLUTO data~\cite{Berger:1983yp} while filled diamonds correspond to 
 TASSO measurements~\cite{Braunschweig:1990yd}. The shaded area represents the uncertainty of our 
 calculation and is computed as explained in  Section~\ref{sec:data}.
}
\label{fig:ptsq}
\end{figure}

Finally, it is useful to test the model of Eqs.~\eqref{eq:powerlaw} and~\eqref{eq:powerlog} for
larger values of $p_\perp$. In fact, 
this is necessary to calculate $\langle p_\perp^2 \rangle$,
since it implies integration over the full range of $p_\perp $. For this, we keep the first
of the constraints~\eqref{eq:mupconstraint}, but free the parameter $\ptatmax$.
First, we perform a fit including data up to $p_\perp=2\,{\rm GeV}$. 
As seen in Figure~\ref{fig:tassofit2} and the last entry of 
Table~\ref{tab:power-law-fit}, the model can successfully accommodate this extended range of $p_\perp$. 
This range is, however, not enough to reproduce to a good accuracy the corresponding TASSO measurements
of $\langle p_\perp^2 \rangle$. Thus, we further extend the range analysis of TASSO data to values up 
to $p_\perp=3\,{\rm GeV}$ and use the resulting minimal parameters to estimate $\langle p_\perp^2 \rangle$. 
Figure~\ref{fig:ptsq} shows our estimate and the data from TASSO and PLUTO.

\begin{table*}
 \centering
\begin{tabular}{|c|c|l|c|}
  \hline
\rule{0pt}{1.2em}
Parametrization & Normalization &  \hspace{1.5cm}parameters & $\chi^2_{pt}$ \\ 
& $N=\{N_{14},N_{22},N_{35},N_{44}\}$ & &  \\ 
\hline
\hline
\rule{0pt}{1.2em}
Power-law - I                        & $N_{14}=2.6\pm0.1$                                        & $\power =\{\power_{14},\power_{22},\power_{35},\power_{44}\}$      &  $\chi^2_{14}=0.35$                                    \rule[-0.5em]{0pt}{1.0em}  \\
$p_\perp\in[0.03 - 1.00]$ GeV        & $N_{22}=3.2\pm0.2$    &                                   &  $\chi^2_{22}=0.30$                                    \rule[-0.5em]{0pt}{1.0em} \\
$14 \times 4$ data point             & $N_{35}=4.0\pm0.1$    & $\power_{14} =3.3\pm0.4$,\,\,\,\,$\power_{22} =2.5\pm0.3$          &  $\chi^2_{35}=0.88$                                    \rule[-0.5em]{0pt}{1.0em} \\ 
                                     & $N_{44}=4.4\pm0.2$                                        & $\power_{35} =2.2\pm0.1$,\,\,\,\,$\power_{44} =2.0\pm0.1$          &  $\chi^2_{14}=0.84$                                    \rule[-0.5em]{0pt}{1.0em}\\            
\hline
\hline
\rule{0pt}{1.2em}
Power-law - II                       &$N_{14}=  2.6\pm0.2$                                       & $\power$ = constant                                                &              \rule[-0.5em]{0pt}{1.0em} \\
$p_\perp\in[0.03 - 1.00]$ GeV        &$N_{22}=  3.3\pm0.2$  &                                                                    &                                                   \rule[-0.5em]{0pt}{1.0em} \\
$56$ data points                     &$N_{35}=  4.0\pm0.1$  & $\power = 2.2 \pm 0.1$                                             &  2.87                                             \rule[-0.5em]{0pt}{1.0em} \\ 
                                     &$N_{44}=  4.2\pm0.2$                                       &                                                                    &              \rule[-0.5em]{0pt}{1.0em} \\ 
\hline  
\hline
\rule{0pt}{1.2em} 
Power-law - III                      &$N_{14}=  2.6\pm0.2$                                          & $\alpha$ = $\power_0+\tilde{\power}\log(Q/Q_0)$                          &            \rule[-0.5em]{0pt}{1.0em}  \\
$p_\perp\in[0.03 - 1.00]$ GeV        &$N_{22}=  3.3\pm0.2$ & $Q_0=14\,{\rm GeV}$                                                &                                                     \rule[-0.5em]{0pt}{1.0em} \\
  $56$ data points                   &$N_{35}=  4.0\pm0.1$ &                                                                    &  0.66                                               \rule[-0.5em]{0pt}{1.0em} \\ 
                                     &$N_{44}=  4.4\pm0.2$                                          &  $\power_0 = 3.1 \pm 0.4$,\,\,\,\,$\tilde{\power} = -1.0 \pm 0.4$        &            \rule[-0.5em]{0pt}{1.0em}\\            
\hline          
\hline
\rule{0pt}{1.2em}
Power-law - IV                       & $N_{14}=2.6\pm0.2$                                           & $\alpha$ = $\power_0+\tilde{\power}\log(Q/Q_0)$                          &            \rule[-0.5em]{0pt}{1.0em}  \\
$p_\perp\in[0.03 - 2.00]$            & $N_{22}=3.2\pm0.3$    & $Q_0=14\,{\rm GeV}$                                                   &                                                \rule[-0.5em]{0pt}{1.0em} \\
$76$ data points                     & $N_{35}=4.0\pm0.1$    & $\power_0 = 3.5 \pm 0.3$,\,\,\,\,$\tilde{\power} = -1.1 \pm 0.3$            &  0.95                                          \rule[-0.5em]{0pt}{1.0em} \\ 
                                     & $N_{44}=4.3\pm0.2$                                                 & $\ptatmax =0.219\pm0.005$                                    &            \rule[-0.5em]{0pt}{1.0em}\\            
\hline  
\hline
\end{tabular}
\caption{Fits of the TASSO four cross sections, corresponding to the power-law parameterization of the $p_\perp$ distributions. 
Parametrization I refers to 4 independent fits (one for each data set corresponding to a different c.m. energy) using the functional 
form of Eq.~\eqref{eq:powerlaw}, with constant $\power$ and $\masspara$ parameters. 
Parametrization II refers to the simultaneous fit of the four data sets, using the functional form of  Eq.~\eqref{eq:powerlaw}, 
with constant $\power$ and $\masspara$ parameters.  
Parametrization III refers to the simultaneous fit of the four data sets, using the functional form of  Eq.~\eqref{eq:powerlaw}, 
with Q-dependent $\power$ and $\masspara$ parameters.}
\label{tab:power-law-fit}
\end{table*}

%%%%%%%%%%%%%%%%%%%%%%%%%%%%%%%%%%%%%%%%%%%%%%%
%-------------------------------------
%-------------------------------------
%-------------------------------------
%-------------------------------------
%-------------------------------------

\section{Final remarks}
\label{sec:final}

TMD FFs embed the essence of hadronization, one of the most important manifestations of QCD 
in the non-perturbative regime. It is therefore important to gather as much information as possible 
on these soft quantities, which cannot be computed, but have to be inferred from experiment.
Over the last few years, several analyses have been performed to extract the {\it polarized} TMD FFs, 
like the Collins function, using the measurements of the Collins asymmetries in $e^+e^- \to h_1 h_2 X$
processes provided by BELLE and BaBar Collaborations, which delivered multidimensional data 
(in bins of $z_1$, $z_2$, $p_{\perp 1}$, $p_{\perp 2}$) with impressive statistics and very high precision.
Unfortunately, no analogous data have been provided on the $p_\perp$-distributions of the 
{\it unpolarized} cross sections, or multiplicities, to allow for the extraction of the unpolarized TMD FFs. 
The absence of these fundamental bricks encumbers the analysis of any other polarized, as well as unpolarized, 
process. 

%While waiting for modern, possibly multidimensional, measurements on the $e^+e^- \to h X$  
%$p_\perp$-distributions of the unpolarized cross sections, we have ``resurrected'' some 
At present, the only available data are some 
old (and almost forgotten) measurements of $e^+e^- \to h X$ cross sections from TASSO and MARKII Collaborations. 
Although these are one-dimensional data 
%(i.e.~$z$-distributions integrated over $p_\perp$ and $p_\perp$-distributions integrated over $z$) 
and are affected by large uncertainties, as discussed in 
Section~\ref{sec:intro}, they have the unique advantage of delivering measurements at different c.m. energies; 
therefore, they can provide a valuable starting point not only to learn about the unpolarized TMD FFs, but also 
to study the physiognomy of their TMD evolution. 

In this article we assess the extent to which the effects of TMD evolution can be observed in these data. 
For this purpose, our main tool is an analysis based on a simple partonic picture in which the cross section 
is factorized, as in 
Refs.~\cite{Anselmino:2005nn,Anselmino:2007fs,Anselmino:2012aa,Anselmino:2013lza,Anselmino:2015sxa}. 
While these types of analysis typically use a Gaussian form, 
we also test a power-law behavior to describe the data, as suggested in Ref.~\cite{Aidala:2014hva}. 
We extend this class of models to the case in which the parameterization is supplemented by some Q dependence, 
and use these results to provide an interpretation within a TMD evolution framework, see Eq.~\eqref{eq:tmdbt}, 
discussing the caveats related to the limited amount of information provided by these data.

We start by modeling the $p_\perp$ dependence of the cross section by a Gaussian shape and fit
the four sets of TASSO cross section data (corresponding to four different c.m. energies) to extract the 
corresponding free parameters, see Table~\ref{tab:gaussian-fit}. Our analysis shows that the Gaussian 
distribution can only describe the data up to $p_\perp \sim 0.5$ GeV, provided 
the cross sections are adjusted with an ad-hoc, additive term $\delta Q$. 
In this region no $Q$-evolution effects can be observed in the Gaussian width of the $p_\perp$-distributions.
The difficulties related to the interpretation of these results, however, leads us to consider a different parameterization.
% For $p_\perp$ values larger than $0.5$ GeV, the Gaussian behavior, as expected, breaks down and a more 
% pronounced $Q$-dependence starts becoming visible. 
%To account for this we have adopted a slightly different functional form of the Gaussian width, allowing 
%it to grow logarithmically with Q, see Eq.~\eqref{eq:widthQdep}, as one could expect from a Gaussian 
%behavior of the non-perturbative $g_K$ function in the impact parameter space. 
%However, even with this added degree of flexibility, the fit does not provide a completely successful 
%description of TASSO data, as reported in the lower panel of Table~\ref{tab:gaussian-fit}. 

We then focus on a power-law parametrization of the $p_\perp$ dependence of the cross section, which 
%fragmentation function. 
%Even using the same number of free parameters, it 
shows to be more appropriate than the Gaussian 
model and provides a successful description of the TASSO $p_\perp$-distributions over a much larger range 
of $p_\perp$ values. As a consistency check, we compare the results of our main fits on TASSO data, reported in the 
last two panels of Table\ref{tab:power-law-fit}, with the MARK II and PLUTO data. We find a reasonable 
agreement of our model in both cases.
% A detailed discussion of how this model can be connected to the framework of TMD 
% factorization is provided in Section~\ref{sec:powerlaw}, where we show that simple arguments based on the 
% asymptotic behavior of the non-perturbative terms of the TMD FF in $b_\perp$ space allow to ...
Finally, we provide an argument to explain that the Q dependence in the power-law can be consistent with a 
logarithmic behavior, in the large $b_\perp$ limit, of the function $g_K$, which encodes the non-perturbative
evolution effects in the definition of the TMD FF.

The nature of these data forces us to be cautious with the interpretation of our results. In fact, 
it seems unlikely that these data by themselves would allow to disentangle the TMD effects from other $Q$-dependence
in the data. This is related to the integration over $z$, which induces a degree of ambiguity 
in the possible interpretations. Thus, one should further test any conjecture 
with multi-dimensional data.

In the foreseeable future, unpolarized single-hadron production at fixed energies by BELLE and BaBar 
Collaborations, differential in both $z$ and $p_\perp$, may indeed provide enough constraints on the 
$z$-dependence of the fragmentation process, allowing for the possibility of a full TMD analysis when 
combined with the TASSO and MARK II data.

%Mention about issues related to uncontrolled normalization

% Finally, we focused on the challenges offered by having to deal with z-integrated data, as z-integration 
% washes out an essential part of information about possible correlations among $z$, $p_\perp$ and $Q$ which 
% would be crucial for a full TMD analysis.

%%%%%%%%%%%%%%%%%%%%%%%%%%%%%%%%%%%%%%%%%%%%%%%%%%%%%
\section*{Acknowledgments}
 We thank Mauro Anselmino for his contributions during the initial stages of this work 
 and for many useful discussions that followed. 
 We are very grateful to Stefano Melis for helping us in finding references to the 
 experimental data used in this paper, and for very helpful discussions and cross-checks.\\
 The work of J.O.G.H was partially supported by Jefferson Science Associates,
 LLC under  U.S. DOE Contract \#DE-AC05-06OR23177 and by U.S. DOE Grant \#DE-FG02-97ER41028.\\
 R.T. wishes to thank A. Mirjalili for his continuos support throughout this project, and 
 Yazd University for financing her stay at the University of Turin, where she performed 
 part of her work.   
%%%%%%%%%%%%%%%%%%%%%%%%%%%%%%%%%%%%%%%%%%%%%%%%%%%%%
%\Urlmuskip=0mu plus 1mu\relax
\bibliographystyle{elsarticle-num}
\bibliography{sample}

%%%%%%%%%%%%%%%%%%%%%%%%%%%%%%%%%%%%%%%%%%%%%%%%%%%%%
\end{document}